\documentclass[reprint, amsmath,amssymb, aps,]{revtex4-2}

\usepackage{amssymb}
\usepackage[caption=false]{subfig}
\usepackage{amssymb}
\usepackage{amsmath}
\usepackage{graphicx}
\usepackage[caption = false]{subfig}
\usepackage{verbatim}
\usepackage{bm}
\usepackage[normalem]{ulem}
\usepackage[utf8]{inputenc}
\usepackage{multirow}
\usepackage{caption}

\begin{document}

\title{Constraining the equation of state in modified gravity via universal relations}

\author{Victor~I.~Danchev$^1$}
\homepage{Electronic address: \url{vidanchev@uni-sofia.bg}}
\author{Daniela~D.~Doneva$^{2,3}$}
\homepage{Electronic address: \url{daniela.doneva@uni-tuebingen.de}}
\affiliation{$^1$Department of Theoretical Physics, Faculty of Physics, Sofia University, Sofia 1164, Bulgaria}
\affiliation{$^2$Theoretical Astrophysics, Eberhard Karls University of Tübingen, Tübingen 72076, Germany}
\affiliation{$^3$INRNE -- Bulgarian Academy of Sciences, 1784 Sofia, Bulgaria}

\date{\today}

\begin{abstract}
Modern multi-messenger astronomical observations and heavy ion experiments provide new insights into the structure of compact objects.
Nevertheless, much ambiguity remains when it comes to super dense matter above the nuclear saturation density such as that found within neutron stars.
This work explores equation of state (EOS)-independent universal relations between the physical parameters of static neutron stars at the maximum-mass point, previously proven for General Relativity (GR) and used to constraint EOS candidates within the GR framework.
We explore 53 different EOS candidates and prove that these relations hold for scalarized neutron stars in massive scalar-tensor theories, exploring also the effect of non-baryonic EOS.
We further show that the relation's fit parameters are highly dependent on the theory's parameters. 
On the basis of these relations multiple constraints on the EOS can be derived and it turns out that they can be significantly different than the GR ones. 
This demonstrates  the importance of taking into account the modified gravity effects even when imposing constraints on the EOS.  
\end{abstract}

\maketitle


\section{Introduction}\label{introduction}
The Scalar-Tensor theories (STT) of gravity are among the earliest considered and most natural extensions of General Relativity (GR) due to their mathematical self-consistency and physical foundation  \cite{Damour:1993hw,Damour:1996,Harada_1,Harada_2,Doneva_2013}.
While weak-field tests can place strong restrictions on alternative theories of gravity, the classes of STT allowing for scalarization predict deviations from GR only in the strong regime.
Due to the existence of ``no hair" \ theorems for the considered class of scalar-tensor theories, however, black holes are ruled out as candidates for scalarization. 
This makes neutron stars the ideal laboratory for the study of STT and other modified theories of gravity in the strong regime \cite{Berti_2015,Doneva:2017jop}.
Since the scalarization of neutron stars is a non-perturbative effect it can lead to significant deviations from GR and observational traits.
While massless STTs lead to the largest deviations in the compact objects' properties, their allowed parameters are significantly constrained by astrophysical observations \cite{Freire_2012,Antoniadis_2013,Shao:2017gwu}. 
The parameters within these constraints do not lead to any significant deviations from GR.
However, this situation is dramatically different when one introduces a scalar field mass. 
This leads to a characteristic lengthscale $\lambda_{\phi} = 2 \pi / m_{\phi} $, that is the Compton wavelength, beyond which effects of the scalar field are suppressed.
This introduces a much wider range of allowed values for the theory parameters and leads to consistency with current experimental evidence while predicting significant deviations from pure GR in the structure of Neutron Stars specifically for low scalar field masses \cite{Ramazanoglu:2016kul,Yazadjiev:2016pcb,Kalin_2018}.

A known obstacle in the study of modified theories of gravity through neutron stars has been the ambiguity in the Equation of State (EOS) governing their internal structure.
While the EOS for nuclear matter bellow the saturation density is well known, the central regions of neutron stars reach higher densities under which the exact state of matter is not yet well understood.
Considerable efforts are invested in the attempts to constraint the EOS via observations \cite{Lattimer:2015nhk,Ozel:2016oaf,Miller:2019cac} and in addition, multiple dimensionless universal relations have been explored which relate different Neutron Star properties to a high degree independently of the EOS \cite{Yagi:2016bkt,Doneva:2017jop}.
The observations may differ from the expected model due to a modified gravitational theory or due to different internal structure of the matter.
 While the studies of the latter case are gaining more and more momentum recently, very little has been done in the direction of exploring how GR modification would alter the existing EOS constraints. 
 For example, the universal relations connecting the  moment of inertia and the neutron star compactness  \cite{Lattimer_1,Breu_1} have been generalized to modified theories of gravity \cite{Doneva_2015,Kalin_1_Uni,Kalin_2_Uni,Popchev_2018}, the neutron star pulse profiles  was examine in the context of STT in \cite{Silva:2018yxz}. 
 A detailed study in the subject, including all possible observations in a variety of theories, is still lacking.

A new type of universal relation in the static case of GR neutron stars was recently explored by D. Ofengeim \cite{Ofengeim_2020} based on an observation of L. Lindblom \cite{Lindblom_2010} that the high-density nuclear matter EOS can be well parametrized by two parameters only.
Considering the maximum mass point of a sequence of neutron stars $M_*$ and the corresponding radius $R_*$, central density $\rho_c$, pressure $P_c$ and speed of sound $c_{sc*}$, it has been shown that certain triplets of these parameters are connected in a way which is independent of the EOS.
More concretely, it has been shown that for a diverse set of very different baryonic EOS in GR, the set of points $(M_*, R_*, x_*)$ form with high a accuracy a single surface, where $x_* = \rho_c, P_c, c_{sc*}$.
As shown by Ofengeim, this type of universality can lead to constraints on the EOS of compact objects in GR based on observational data and physical considerations.

In this work, we extend the study of this universality to the case of scalar-tensor theories of gravity and further consider the addition of non-baryonic EOS.
We confirm that a class of massive scalar-tensor theories respects this form of maximum-mass universality and explore its dependence on the introduction of non-baryonic EOS.
The obtained relations can be used to explicitly demonstrate that the existing EOS constraints considered in \cite{Ofengeim_2020} are significantly altered by the GR modifications. In addition, with the help of these universal relations one can constraining the parameters of the alternative theories of gravity independently of the EOS.
Another important goal of the present studies is also to see up to what extend the EOS constraints derived from universality are sensitive to the EOS used to obtain the relations even when taking into account the most basic EOS properties.

The work is structured as follows. We outline the basic equations of massive STT in section \ref{theory} and introduce the proposed maximum mass universality in section \ref{Theory_Uni_Rels}. 
We then present our numerical results in section \ref{Results} and outline the constraints on EOS parameters for different STT parameters in section \ref{Constraints}.
The work ends with Conclusions.

\section{Theory}\label{theory}
We work within the framework of the scalar-tensor theories (STT) with an action in the Einstein frame given by
\begin{eqnarray}\label{action}
S = && \frac{1}{16 \pi G_*} \int d^4 x \sqrt{ - g}( R - 2 g^{\mu \nu} \partial_{\mu} \varphi \partial_{\nu} \varphi - V(\varphi) ) + \nonumber \\
 && S_m( A^2( \varphi ) g_{\mu \nu} , \Phi_m ),
\end{eqnarray}
where $R$ is the Ricci scalar with respect to the Einstein frame metric $g_{\mu \nu}$, $\varphi$ is the scalar field and $ S_m$ is the action of the matter. 
We work in natural units of $c = \hbar = 1$ for simplicity of the notation.
The theory is specified by the conformal factor $A(\varphi)$ and the scalar field potential $V(\phi)$. 
Following \cite{Damour:1993hw}, we use an exponential function, characterized by a single parameter $\beta$, of the form
\begin{eqnarray}\label{coupling_function}
A(\varphi) = e^{ \frac{1}{2} \beta \varphi^2 }.
\end{eqnarray}
In addition, we employ the simplest and most natural $\mathbb{Z}_2$ symmetric scalar potential, namely
\begin{eqnarray}\label{potential}
V(\varphi) = 2 m_{\varphi}^2 \varphi^2,
\end{eqnarray}
where $m_{\varphi}$ is the mass of the scalar field.
This choice of the conformal factor $A(\varphi)$ leads to a STT that is perturbatively equivalent to GR in the weak-field regime but where new effects can appear for strong fields, such as
neutron star scalarization that is considered in the present paper.

The calculations in the present paper are performed in the Einstein frame due to its simplicity while the final results are transformed to the physical Jordan frame. Unless otherwise specified, all quantities with a tilde will be in the Jordan frame and without a tilde -- in the Einstein frame. A thorough discussion about the two frames in the context of scalarized neutron stars can be found e.g. in \cite{Doneva:2013qva} while here we will comment only on the most important points.  The Jordan (physical) frame metric is connected to the Einstein one through the conformal factor $A(\varphi)$ in the following way $\tilde{g}_{\mu \nu} = A^2(\varphi) g_{\mu \nu}$.
The relation between the energy-momentum tensor in the two frames is given by $T_{\mu \nu} = A^2(\varphi) \tilde{T}_{\mu \nu}$.
This leads to the following relations between the energy density and the pressure of an ideal fluid $\rho = A^4(\varphi) \tilde{\rho}$ and $p = A^4(\varphi) \tilde{p}$.

In this paper we are using the standard ansatz for a spherically-symmetric static space-time
\begin{eqnarray}\label{metric}
ds^2 = - e^{ 2\Gamma } dt^2 + e^{ 2 \Lambda }dr^2 + r^2( d\theta^2 + \sin^2{\theta} d\varphi^2 ).
\end{eqnarray}
The dimensionally reduced Einstein frame field equations obtained from the action \eqref{action} with the metric \eqref{metric} are
\begin{widetext}
\begin{eqnarray}\label{Field_Eqs1}
\frac{2}{r}e^{-2\Lambda} \frac{d \Lambda }{dr} & = & 8 \pi G_{*} A^4 \tilde{\rho} - \frac{1}{r^2}(1 - e^{-2\Lambda}) + e^{-2\Lambda } \left( \frac{d \varphi}{dr} \right)^2 + \frac{1}{2} V(\varphi) \\ \label{Field_Eqs2}
\frac{2}{r}e^{-2\Lambda} \frac{d \Gamma }{dr} & = & 8 \pi G_{*} A^4 \tilde{p} - \frac{1}{r^2}(1 - e^{-2\Gamma} ) + e^{-2\Lambda} \left( \frac{d \varphi}{dr} \right)^2 - \frac{1}{2} V(\varphi) \\ \label{Field_Eqs3}
\frac{d\tilde{p}}{dr} & = & ( \tilde{p} + \tilde{\rho} ) \left( \frac{d\Gamma}{dr} + \frac{d \ln{A(\varphi)} }{d\varphi} \frac{d\varphi}{dr} \right) \\ \label{Field_Eqs4}
\frac{d^2 \varphi}{dr^2} & = &  \left( \frac{d\Lambda}{dr} - \frac{d\Gamma}{dr} - \frac{2}{r} \right) \frac{d\varphi}{dr} + 4 \pi G_{*}A^4(\varphi)( \tilde{\rho} - 3 \tilde{p} ) e^{2\Lambda} \frac{d \ln{A(\varphi)}}{d\varphi} + \frac{1}{4} \frac{dV(\varphi)}{d\varphi} e^{2 \Lambda}.
\end{eqnarray}
\end{widetext}

Naturally the system of equations \eqref{Field_Eqs1}--\eqref{Field_Eqs4} is incomplete and must be supplemented by an Equation of State (EOS) for the neutron star matter.
The complete system for each of the EOS choices can be solved by setting the appropriate boundary conditions at the centre of the star and at spatial infinity.
The central values are given by $\tilde{\rho}(0) = \tilde{\rho}_c $, $\Lambda(0) = 0$ and $ \frac{d\varphi}{dr} = 0$.
On the other hand, from the asymptotic flatness at infinity we have $\Gamma( \infty ) = 0$, $\Lambda( \infty ) = 0$ and $\varphi( \infty ) = 0$.
The central value $\varphi( 0 ) = \varphi_c$ is determined through a shooting method by demanding that the asymptotic value of $\varphi$ tends to zero.
Following an analogous procedure to that in GR, the surface of the star in the Einstein frame $r_S$ is determined by the condition $p(r_S) = 0$.
The physical surface radius in the Jordan frame which is used for all of the fits and plots further in this work is found as $\tilde{r}_S = A(\varphi( r_S ) ) r_S $.

As previously discussed in \cite{Ramazanoglu:2016kul,Yazadjiev:2016pcb}, the addition of the scalar field mass to the theory can reconcile values of $\beta$ which would otherwise predict scalar modes in the gravitational wave emissions unobserved for binary systems of compact objects \cite{Freire_2012,Antoniadis_2013}. 
In order for this to happen, the mass of the scalar field should be chosen in such a way that  its Compton wavelength is much smaller than these binaries' separation $\lambda_{\varphi} << r_{\mathrm{binary}}$.
Furthermore, as shown in \cite{Kalin_2018}, the scalar field mass suppresses the scalar field and with the increase of $m_\varphi$ the scalarized solutions approach the GR results more or less monotonically.
Therefore,  the highest deviation from GR can be expected in the $m_\varphi \rightarrow 0$ (massless) limit.

\begin{figure}[!h]
\centering
\includegraphics[scale=0.5]{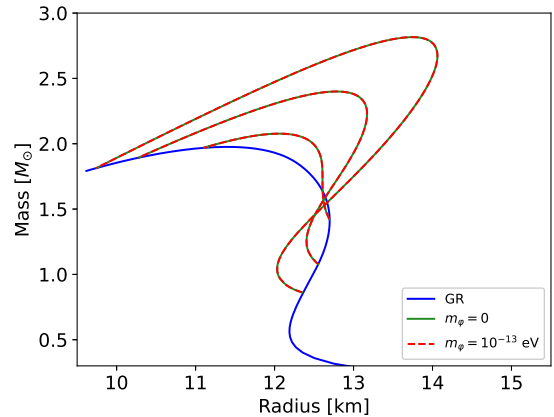}
\caption{Mass-Radius relations of the ALF2 EOS for GR and $\beta = -5, -6, -7$. Massless and $m_{\varphi} = 10^{-13}$ eV cases for each $\beta$ value are shown in solid green and dashed red lines respectively.}\label{ALF2_Example}
\end{figure}

In order for the theory to not to predict any influence on the binary pulsar observation, the Compton wavelength should be smaller that the orbital separation between the two compact objects.
The largest constraint on massless STT comes from PSR J0348+0432.
The measured separation is $r_{\mathrm{binary}} \sim 10^9$ m and the typical size of a neutron star is on the order of $r_{\mathrm{NS}} \sim 10^4 $ m. 
Therefore, a scalar field mass of $m_{\varphi} = 10^{-13} $ eV will leave $\beta$ practically unconstrained from the binary pulsar observation as its Compton wavelength corresponds to $\lambda_{\varphi} \sim 10^7 $ m.
On the other hand, such a low mass' effect is virtually indistinguishable from the massless case as demonstrated in Fig.\ref{ALF2_Example} for the ALF2 EOS.
The ALF2 is a hybrid EOS consisting of mixed APR nuclear matter and color-flavor-locked quark matter.
We have observed the same behavior for other EOS (both baryonic and non-baryonic). Therefore, for the actual computations this low mass does not produce any significant deviation from the massless case, while it allows us to work with low $\beta$ values which produce large differences with respect to pure GR.

Numerically, the relative differences in the neutron star masses and the neutron star radii between the cases $m_{\varphi} = 0$ and $m_{\varphi} = 10^{-13}$ eV respectively does not exceed $10^{-2}$ as obtained for several very different EOSs.
We have therefore proceeded with the numerical computation of all models in the massless framework.
Indeed, this approximation's validity is supported by the results since the minor corrections to the fits obtained from the mass $m_{\varphi} = 10^{-13}$ eV are small compared to the obtained fits rms.

\section{Maximum-Mass Universal properties}\label{Theory_Uni_Rels}
Lindblom \cite{Lindblom_2010} demonstrated that for a wide class of EOS models the $P(\rho)$ dependence in the neutron stars core can be well approximated with only two parameters. 
Following this idea, Ofengeim showed that within the framework of General Relativity (GR), the sets of points $( M_{*}, R_{*}, x_{c*} )$ of maximum-mass neutron stars belong with high accuracy to a single 2-dimensional surface \cite{Ofengeim_2020} and can be further fitted by analytic expressions.
In this notation, $M_*$ and $R_*$ signify the mass and radius at the maximum-mass point of a sequence of neutron star models while $x_* = \tilde{\rho}_{c*}, \tilde{p}_{c*}, \tilde{c}_{sc*}$ takes the values of central density, pressure and speed of sound for that star.
It was demonstrated that the combination of these analytical fits together with certain constraints on $M_*$, $R_*$ and their relations to $x_*$ can lead to novel restrictions on the properties of neutron star matter in GR.
Following his work, we explore universality in the functional dependence of the central pressure, density and speed of sound as functions of the neutron star's radius and mass at the maximum-mass point outside of GR.
In this work we confirm that this form of universality also holds for scalar-tensor theories of the type \eqref{action} for a wide range of physically meaningful $\beta$ parameter but the differences in the fitting parameters and the qualitative behavior can be large.
In order to verify this universality and obtain quantitative results of its accuracy, we have used the same fit as in the original work
\begin{eqnarray}\label{Primary_Fit}
x_* = x_0 \left( \frac{a_x}{R^*_{\mathrm{mix}, x} - b_x} \right)^{\mathfrak{p}_x},
\end{eqnarray}
where we have made a slight change of notations and used the definition
\begin{eqnarray}\label{Mix_Angle}
R^*_{\mathrm{mix}, x} = R_{*} \cos{\phi_x} + r_{g*} \sin{ \phi_x }.
\end{eqnarray}
$\phi_x$ has the interpretation of a mixing angle as in the original work \cite{Ofengeim_2020}.
This angle can be interpreted as the rotational angle between the $R_*$ axis and an approximate line obtained through the projection of the $x_*$'s cross-section with a horizontal plane.
The parameter $r_{g*}$ is simply the star's Schwarzschild radius at the maximum mass point $r_{g*} = 2 G_* M_*/c^2$.
The four values $a_x, \phi_x, b_x $ and $\mathfrak{p}_x$ are the fit parameters which are varied in order to minimize \eqref{Primary_Fit}'s rms while the values $x_0$ carry the appropriate dimensionality.
This dimensionality has been set by the saturation density of symmetric nuclear matter and naturally -- the speed of light.
We have chosen $\rho_0 = 2.8 \times 10^{14} \ \mathrm{g \, cm}^{-3}$, $P_0 = \rho_0 c^2 $ and $ c_0 = c $.
In addition to extending the GR universal relations to scalar-tensor theories, we further considered non-baryonic EOS in order to be able to compare the deviations coming from modifying the theory of gravity and from considering a wide set of EOS.
The following section \ref{Results} presents our results for these fits which confirm that the universality holds outside of GR with accuracy of the same order as that in GR.
Additionally, the results demonstrate that the universality worsens significantly with the addition of non-baryonic EOS (specifically with strong phase transitions) in all theories.

Considering $\rho_{c*}( M_* , R_* )$ and $P_{c*}( M_* , R_* )$ as a system of equations, the dependence can be easily inverted.
Substituting these results in the fit for $c_{sc*}$, one can obtain a secondary fit, entirely dependent on the original parameters as
\begin{eqnarray}\label{Secondary_Fit}
y_* = && y_0 \left\{ A_y \left[ ( \rho_0/\tilde{\rho}_{c*} )^{1/\mathfrak{p}_{\tilde{\rho}}} \cos{\Phi_y} + \right. \right. \nolinebreak\\
&& \left. \left. (\rho_0 c^2/\tilde{p}_{c*} )^{1/\mathfrak{p}_{\tilde{p}}} \sin{\Phi_y} - B_y \right] \right\}^{q_y},
\end{eqnarray}
where $y_* = M_*, R_*, \tilde{c}_{sc*}$.
One can easily show that $q_{\tilde{p}_{c*}} = q_{\tilde{\rho}_{c*}} = 1$ and $q_{\tilde{c_{sc*}}} = - \mathfrak{p}_{\tilde{c}}$.
The remaining parameters are entirely derived from the previous fits with the appropriate normalization condition.
Following the example with GR \cite{Ofengeim_2020}, a new mixing variable has once again been defined as
\begin{eqnarray}\label{Mix_Second}
\chi_{\mathrm{mix}, y} = \left( \frac{\tilde{\rho}_0}{\tilde{\rho}_{c*}} \right)^{\frac{1}{\mathfrak{p}_{\tilde{\rho}}} } \cos{ \Phi_y } + \left( \frac{\tilde{\rho}_0 c^2}{\tilde{P}_{c*}} \right)^{ \frac{1}{\mathfrak{p}_{\tilde{P}}}} \sin{ \Phi_y }.
\end{eqnarray}
We verify in section \ref{Results} that these secondary fits are up to a large extend independent of the EOS for the STT as they are for GR, and they are significantly influenced by non-baryonic EOS.

All of these relations and their fit errors allow us to explore several constraints in Sec. \ref{Constraints}, based on causality and the highest measured neutron star masses.
Unlike the analogous constraints in GR, these can be interpreted both as EOS and theory constraints based on observational data as is further commented in the following sections.

\section{Numerical Results}\label{Results}
We have explored a total of 53 EOS including a diverse range of phases and models.
24 of these are piecewise polytropic approximations based on \cite{APR4_poly} including SLy, APR1-4, MPA1, MS1-2, MS1b, H1-H7, WFF1-3, ENG and ALF1-4.
The remaining 29 are tabulated EOS obtained from the CompStar Online Supernovae Equations of State (CompOSE) including GM1\_Y4-6 \cite{GM_eos_1,GM_eos_2,Oertel_2015}, QHC18-19 \cite{QHC_eos_1,QHC_eos_2,QHC_eos_3,QHC_eos_4}, SK255, SK272, SKa/b \cite{Oertel_2015,SK1_eos_1, SK1_eos_2,SK1_eos_3}, SkI2-6, SkMp, SkOp \cite{Oertel_2015,SK1_eos_3,SK2_eos_1,SK2_eos_2}, SLy2,4,9,230a \cite{Oertel_2015,SK1_eos_3,SLY_eos_1,SLY_eos_2,SLY_eos_3} and others \cite{SK1_eos_3,others_eos_1,others_eos_2,others_eos_3,others_eos_4,others_eos_5,others_eos_6}.
For each of the 53 EOS, we have computed a sequence of models based on adaptive-step Dormand-Prince Algorithm \cite{Dormand_Prince} and numerically found the maximum mass point.
We have thus obtained the parameters $P_{c*}$, $\rho_{c*}$, $c_{sc*}$, $M_*$ and $R_*$ for GR and five different values of the coupling $\beta$ from \eqref{coupling_function}.
The actual sequence of values used is $\beta = \{ -5 , -5.5 , -6 , -6.5 , -7 \}$.
All of these models have been computed for a scalar field with negligible mass, considering the fact that the scalar field mass on the order of $m_{\varphi} \sim 10^{-13} $ eV produces results virtually indistinguishable from these for the massless case as seen on Fig. \ref{ALF2_Example} and supported by the rms values.

\begin{figure*}[!h]
\centering
\subfloat[Pressure $P_{c*}$ fits]{\includegraphics[width = 6cm]{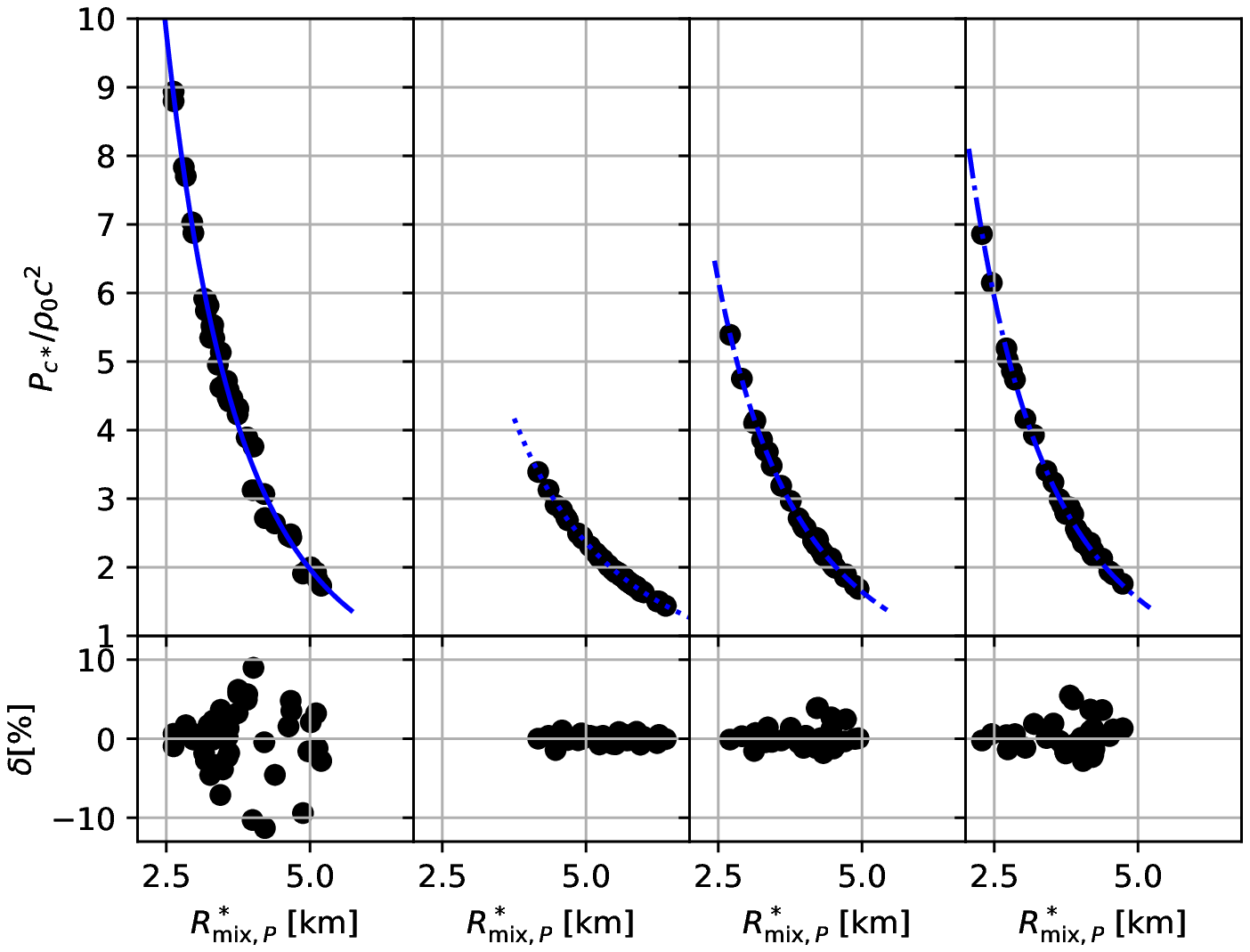}}
\subfloat[Density $\rho_{c*}$ fits]{\includegraphics[width = 6cm]{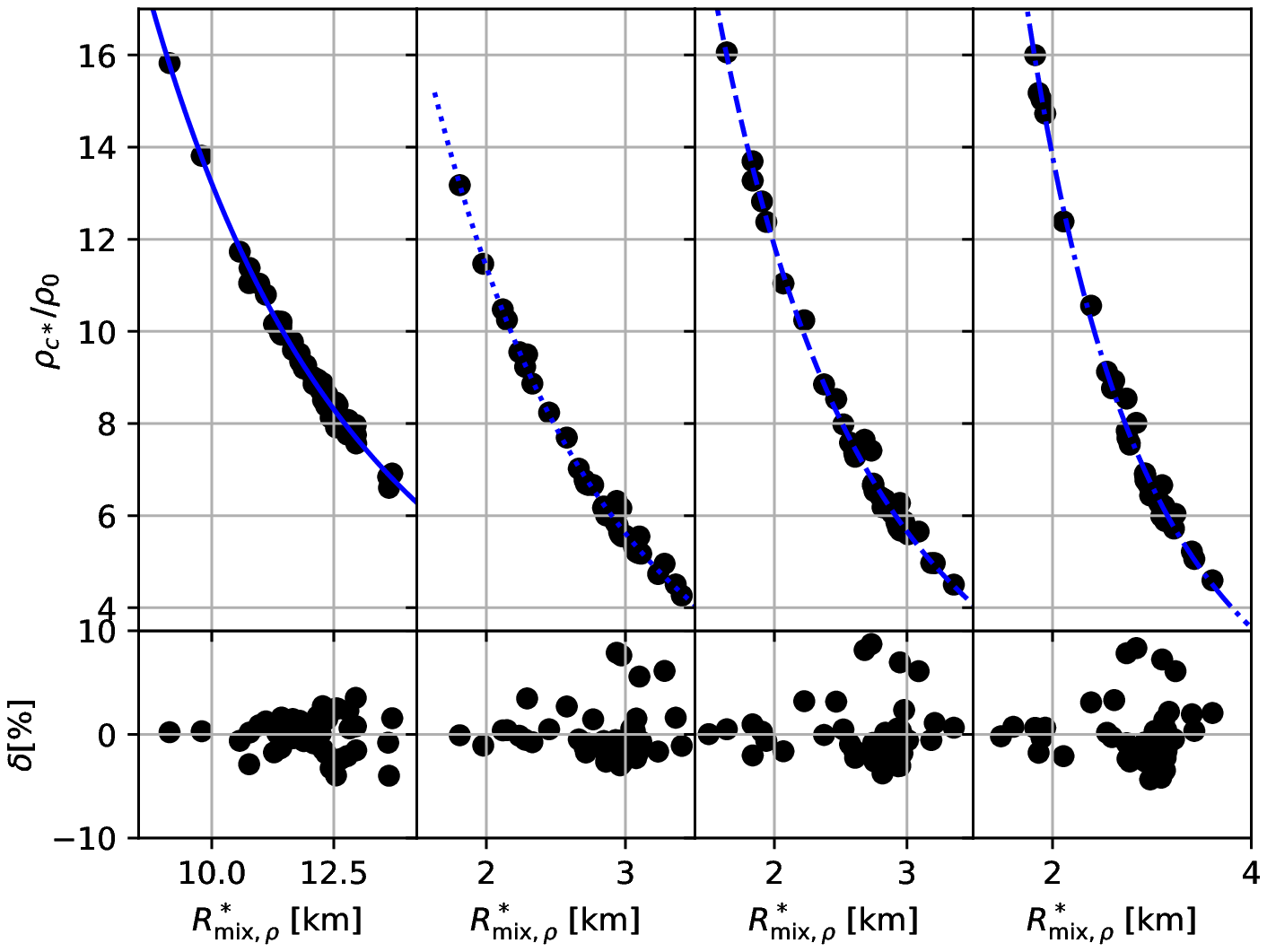}}
\subfloat[Speed of sound $c_{\mathrm{cs}*}$ fits]{\includegraphics[width = 6cm]{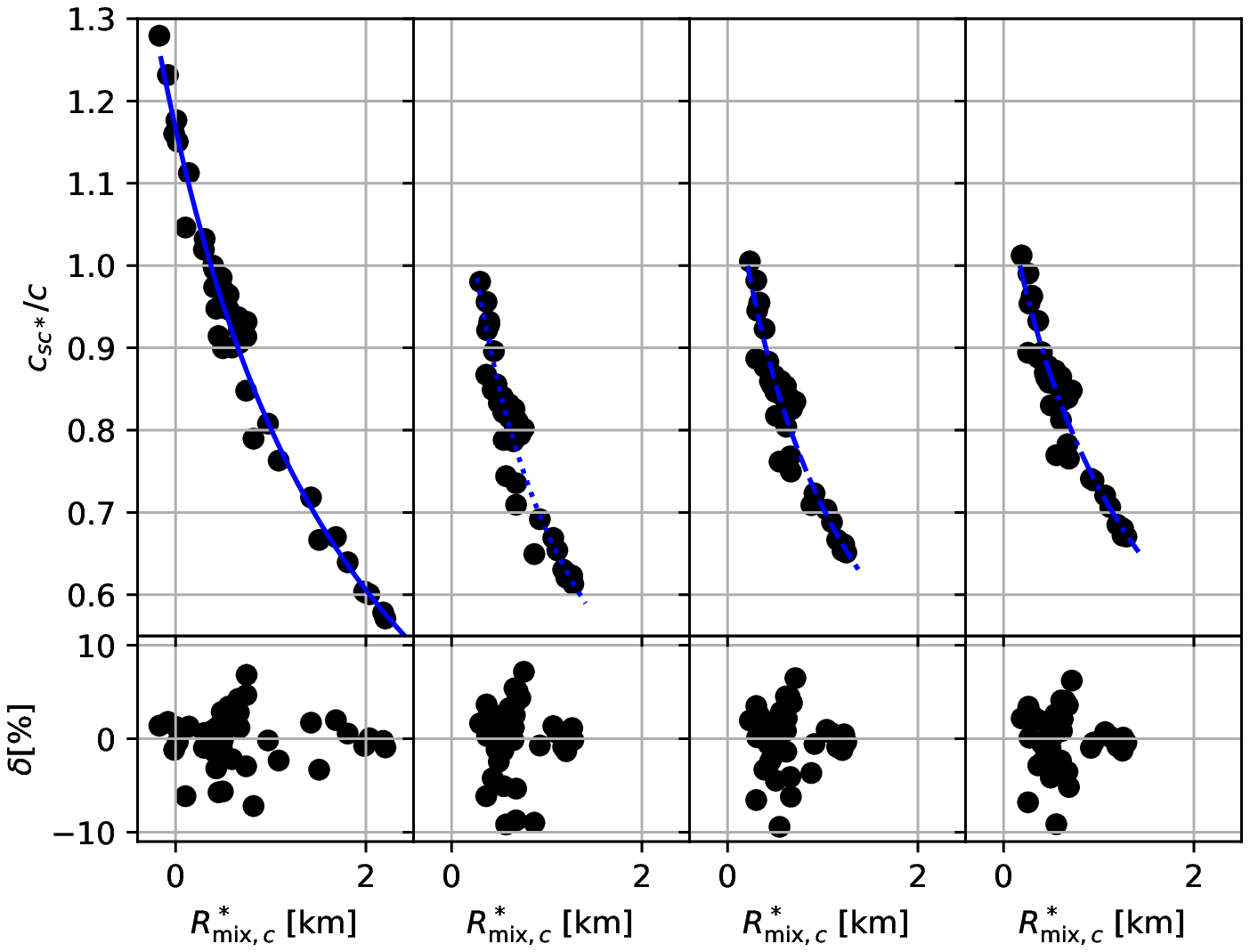}}\\
\caption{$x_*(R^*_{\mathrm{mix},x_*})$ fits of the form \eqref{Primary_Fit} going from left to right: GR (solid line), $\beta = -5$ (dotted line), $\beta = -6$ (dashed line) and $\beta = -7$ (dash-dotted line) for the scalar-tensor theory.}\label{Primary_Fits_Fig}
\end{figure*}

Figure \ref{Primary_Fits_Fig} demonstrates the primary fits \eqref{Primary_Fit} obtained with all baryonic EOS for GR and three of the theories with integer $\beta$.
We have left only these for better readability of the plots.

It is apparent from the relative errors that the fit accuracy is at least as good for each of the scalar-tensor theory parameters as it is for GR.
The actual rms and maximum deviations for each of the fits can be found in Table \ref{Primary_Fit_Table}.
A promising result apparent from the table, is that the fit parameters are significantly different for the different $\beta$ values.
Although the universality is present for generalized theories, its quantitative aspects are not theory-independent and offer significant distinction between GR and alternative theories on one hand and the parameters of a single theory on the other.

\begin{figure*}[!h]
\begin{center}
\includegraphics[width = 9cm]{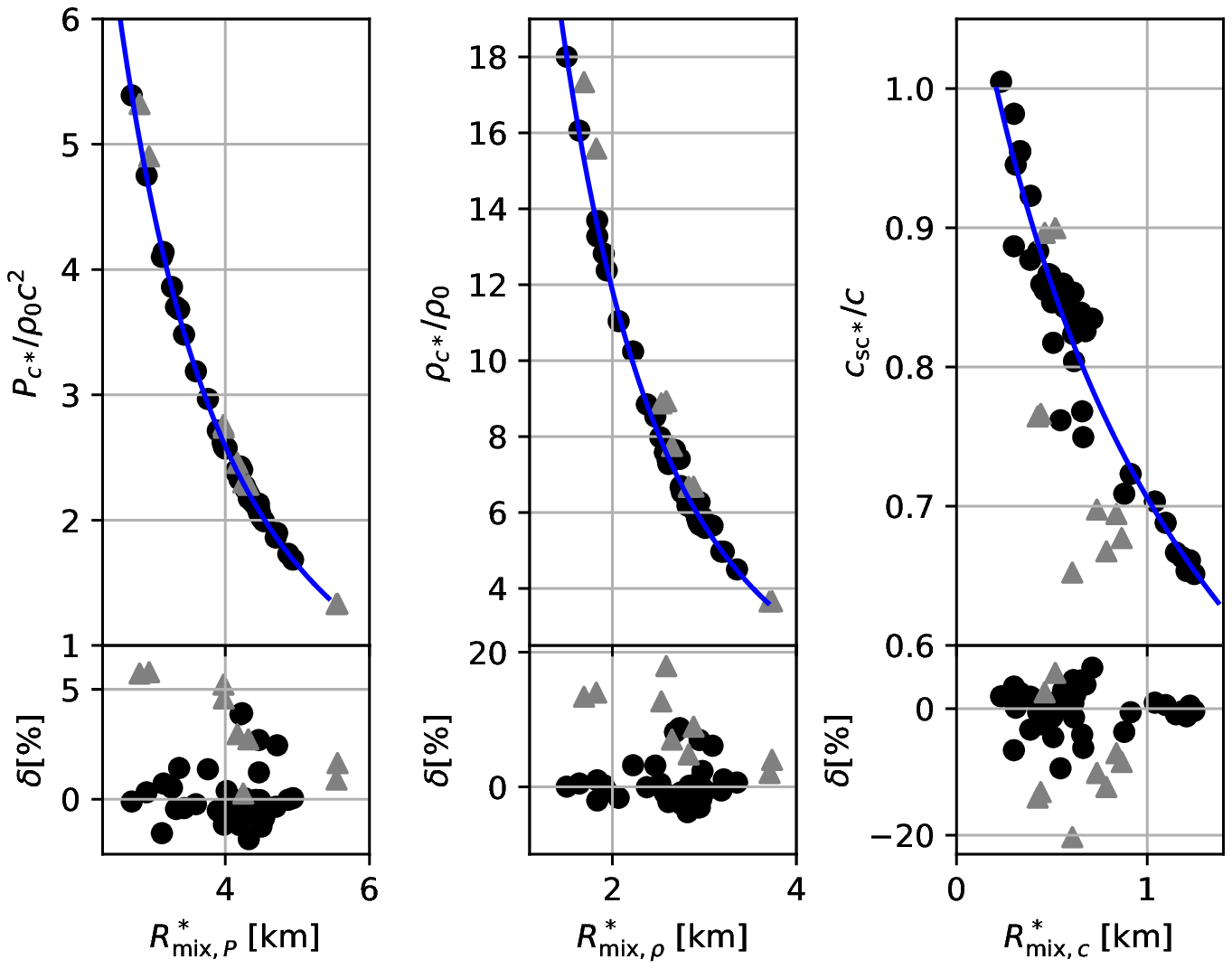}
\caption{$x_*(R^*_{\mathrm{mix},x_*})$ fits of the form \eqref{Primary_Fit} for all EOS at $\beta = -6$. Baryonic points are shown as black circles and non-baryonic points are shown as gray triangles.}\label{Baryonic_Comparison}
\end{center}
\end{figure*}

We have performed a separate set of fits with all the EOS (including non-baryonic and strong phase transitions) which are also available in Table \ref{Primary_Fit_Table}.
It is apparent from the fit coefficients that the universality worsens significantly with the addition of the non-baryonic matter.
This is particularly true for the speed of sound which leads to an even larger worsening of the secondary fits visible in Table \ref{Secondary_Fit_Table}.
For better readability we are giving a single illustrative example of this effect for the case $\beta = -6$, where the complete set of EOS are shown with a distinction between the baryonic and non-baryonic EOS.
Fig. \ref{Baryonic_Comparison} shows all three of the primary fits in this case.
Note that the x-axis $R_{\mathrm{mix},x}^*$ depends on the fit parameters themselves and so we have computed all points' x-value with the same fit coefficients (those for baryons-only).

The following Fig \ref{Secondary_Fits_Fig} demonstrates the derived fits for the same theories as Fig. \ref{Primary_Fits_Fig} for the baryonic-only case, following the form \eqref{Secondary_Fit} with mixing variable \eqref{Mix_Second}.
Once again, the relative fit error for GR and the alternative theories is on the same order demonstrating the validity of the universal relations.


\begin{figure*}[!h]
\centering
\subfloat[Mass $M_*$ fits]{\includegraphics[width = 6cm]{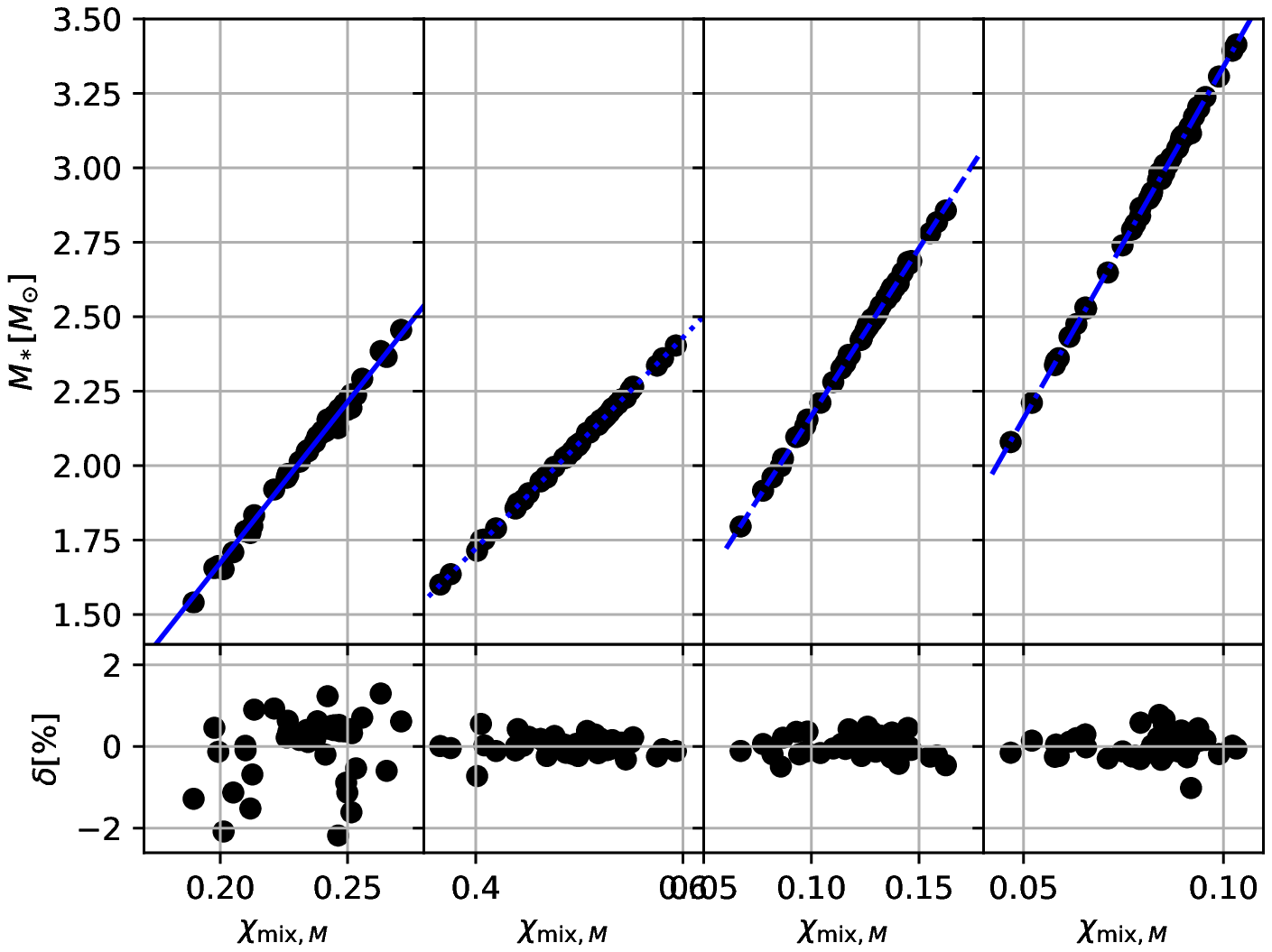}}
\subfloat[Radius $R_*$ fits]{\includegraphics[width = 6cm]{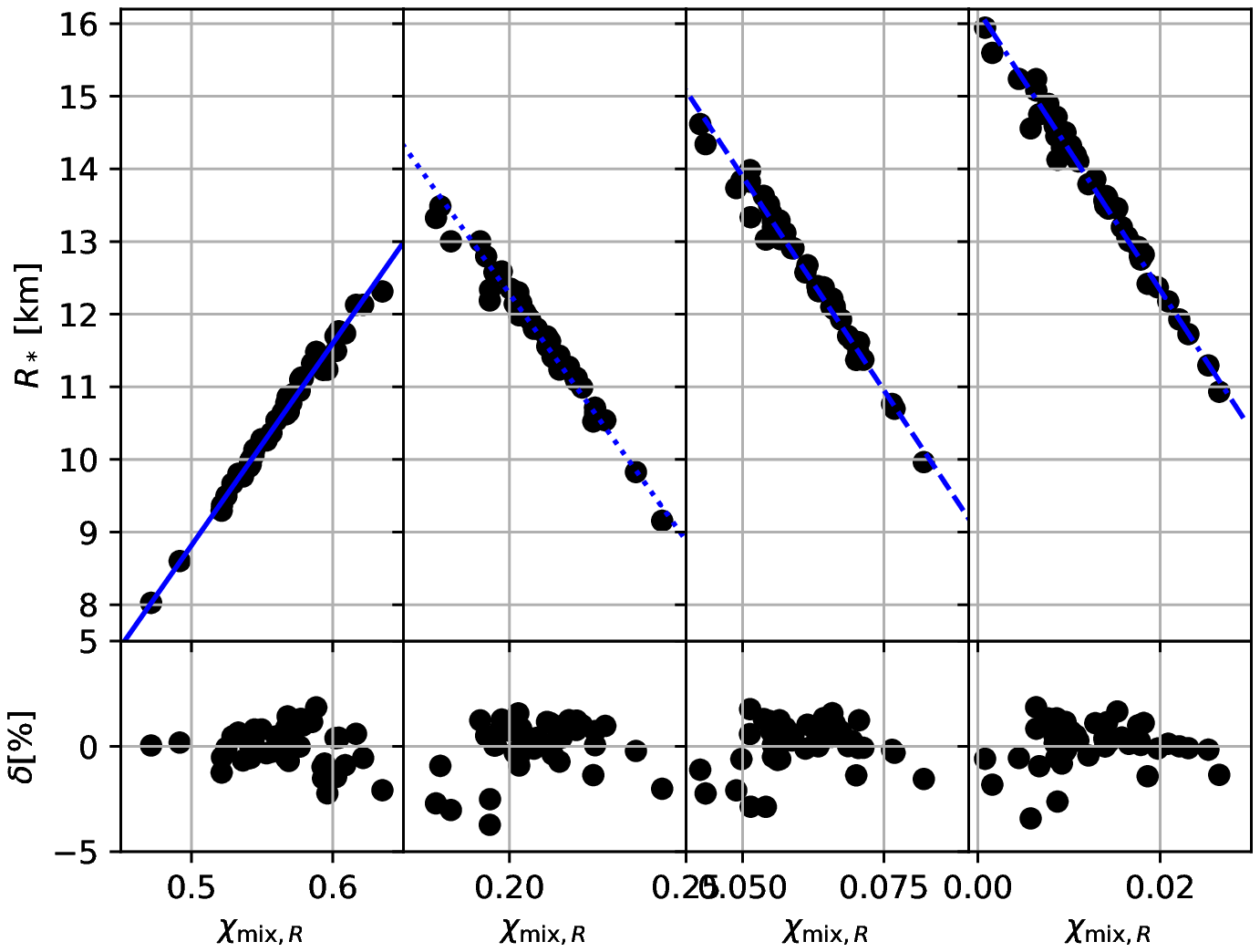}}
\subfloat[Speed of sound $c_{\mathrm{cs}*}$ fits]{\includegraphics[width = 6cm]{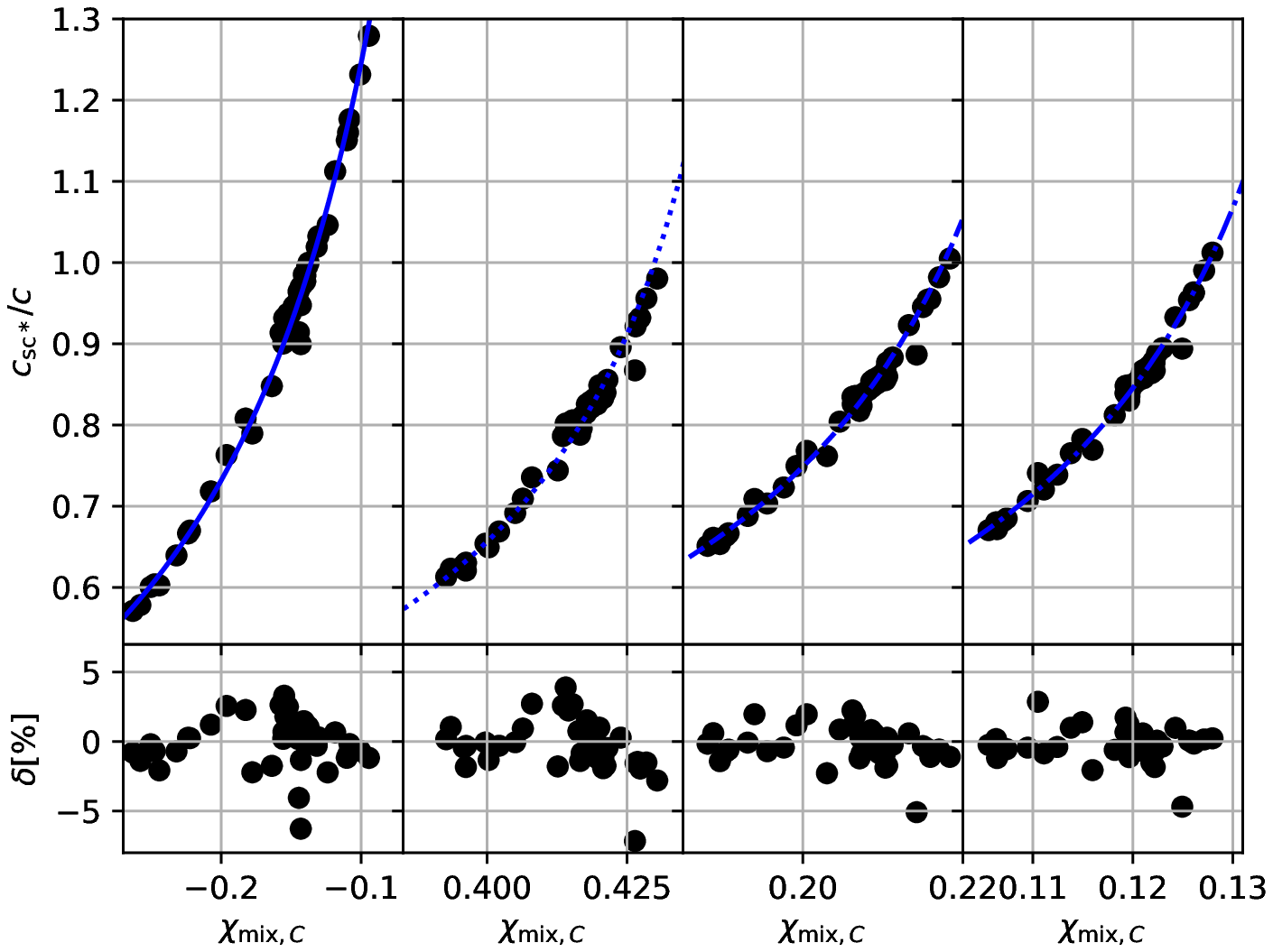}}\\
\caption{$y_*(\chi_{\mathrm{mix},y_*)}$ fits of the form \eqref{Secondary_Fit} going from left to right: GR (solid line), $\beta = -5$ (dotted line), $\beta = -6$ (dashed line) and $\beta = -7$ (dash-dotted line) for the scalar-tensor theory.}\label{Secondary_Fits_Fig}
\end{figure*}

It is apparent from Table \ref{Secondary_Fit_Table}  that the baryonic-only secondary fits are once again much better in particular for the speed of sound $c_{sc*}$.
The highest deviation from the fit increases on average 4--5 times after adding the non-baryonic EOS.
The same occurs in the GR fits as can be seen on the same table and so the effect is not caused by the scalar-tensor theory's specifics.
For all theories and parameters, QHC18 and QHC19 possess the highest deviation from the fit at nearly 25 \% relative difference for the speed of sound.
Removing these and the remaining 5 non-baryonic EOS improves the fits significantly, dropping the highest deviation for the speed of sound to values on the order of 5 \%.
While the improvement in the fit of $M_*$ and $R_*$ is not as significant once non-baryonic EOS are removed, it is still present.
This clearly shows that the universality's accuracy is strongly dependent on the nature of the EOS employed.
In particular, it shows that even a small amount of non-baryonic EOS with strong phase transitions lead to a large reduction in the fit's accuracy.

This reduction in accuracy for a more broad class of EOS does not lead to significant changes in the actual fit parameter values but only their rms and maximum deviation values as seen in both Tables \ref{Primary_Fit_Table}, \ref{Secondary_Fit_Table}. 
This implies that even with a broader set of EOS, the maximum mass universality can be used to reliably tell apart different theories of Gravity.
Indeed, the major impact of using baryonic-only instead of more broad EOS is the more strict constraints for any given gravitational theory as is shown in the next section.

\section{Constraining alternative theories of Gravity through maximum mass Neutron Stars}\label{Constraints}

The aim of the present section is to demonstrate how certain restrictions between neutron star properties or their measured value can be translated into constraints on the EOS and in some cases can even be used to test the theory of gravity in an EOS independent way.
Most of the restrictions used in \cite{Ofengeim_2020} are not directly applicable in our case, though, because they are derived strictly in Einstein's theory of gravity.
The generalization of some of these restrictions to alternative theories, including STT, is not done yet. 
The work on the generalizations of the relations, connected to the neutron star mergers and the maximum compactness of neutrons stars is underway, but these are quite involved and beyond the scope of the present paper. 
That is why we will focus our investigation on two simpler and straightforward constraints, namely causality and maximum mass, in order to demonstrate our idea.
Each of these restrictions can be plotted either in the $M_*$--$R_*$ or in the $\rho_{c*}$--$P_{c*}$ plane considering the fits for the corresponding quantity with their respective maximum error for each theory parameter.

In the case of causality, we must plot the conditions $c_{sc*}( 1 \pm \epsilon_{f,c} ) = 1$ where $f = \{ MR, P\rho \}$ and so $\epsilon_{f,c}$ corresponds to the relative error in the fits $c_{sc*} = c_{sc*}(M_*,R_*)$ and $c_{sc*} = c_{sc*}(\rho_*,P_*)$ for these two values respectively.
Considering the fit definitions \eqref{Primary_Fit} and \eqref{Secondary_Fit}, it is easy to show that the condition $c_{sc*} = 1$ corresponds to the region
\begin{eqnarray}
\left(\frac{\rho_0 c^2}{P_{c*}} \right)^{\frac{1}{\mathfrak{p}_P}} & + \cot{\Phi_c} \left(\frac{\rho_0}{\rho_{c*}} \right)^{\frac{1}{\mathfrak{p}_{\rho}}} =  \nonumber\\
&\frac{1}{\sin{\Phi_c}}\left[ B_c + \frac{(1 \pm \epsilon_{P\rho,c})^{-\frac{1}{q_y}}}{A_c} \right]
\end{eqnarray}
in the $\rho_{c*}$--$P_{c*}$ plane and the region between the two lines
\begin{eqnarray}
M_* + R_* \cot{\phi_c} = \frac{1}{\sin{\phi_c}}\left[ b_c + a_c ( 1 \pm \epsilon_{MR,c} )^{\frac{1}{\mathfrak{p}_c}} \right]
\end{eqnarray}
in the $M_*$--$R_*$ plane.

We can apply directly the observational constraint on the maximum static neutron star mass $ M_* > 1.97 M_{\odot}$ to the $M_*$--$R_*$ plane.
In a manner similar to that for the causality, we can also translate this condition in the $\rho_{c*}$--$P_{c*}$ by considering the fit $M_* = M_*(\rho_*,P_*)$ with a given maximum error $\epsilon_{P\rho,M}$.
The condition for some fixed mass $M_* = \bar{M} = \mathrm{const}$ is then translated to $M_*(\rho_*,P_*)( 1 \pm \epsilon_{P\rho,M} ) = \bar{M}$.
This is once again easily inverted from \eqref{Secondary_Fit} as
\begin{eqnarray}
P_{c*} = \left[ \frac{1}{\sin{\Phi_M}}\left( \frac{\bar{M}^{\frac{1}{q_{M}}}}{A_M} + B_M \right) - \left( \frac{\rho_0}{\rho_{c*}} \right)^{\frac{1}{\mathfrak{p}_{\rho}}} \cot{ \Phi_M } \right]^{-\mathfrak{p}_P}.
\end{eqnarray}

\begin{figure*}[!h]
\subfloat[GR restrictions in $P(\rho)$ plane]{\includegraphics[width = 9cm]{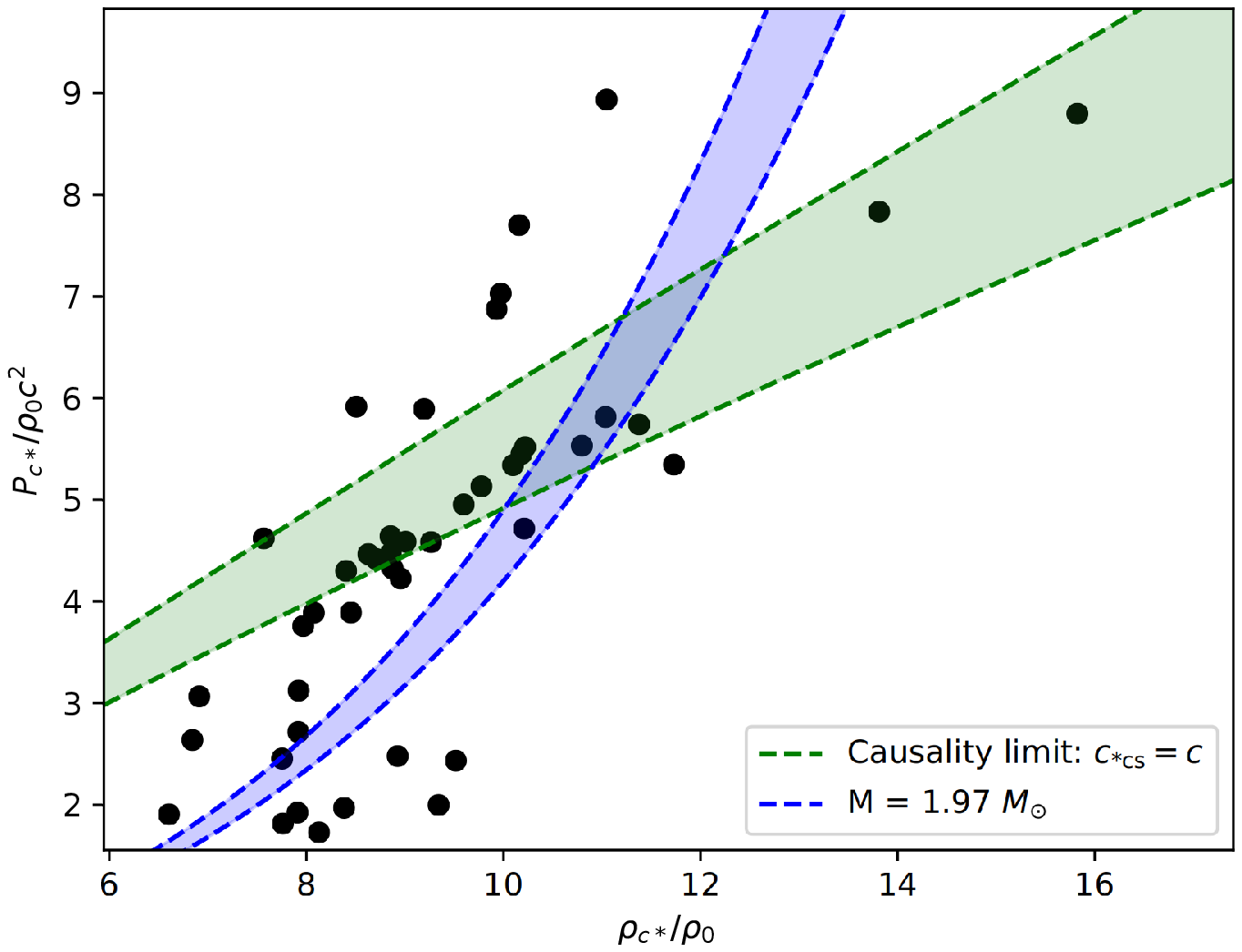}}
\subfloat[GR restrictions in $M(R)$ plane]{\includegraphics[width = 9cm]{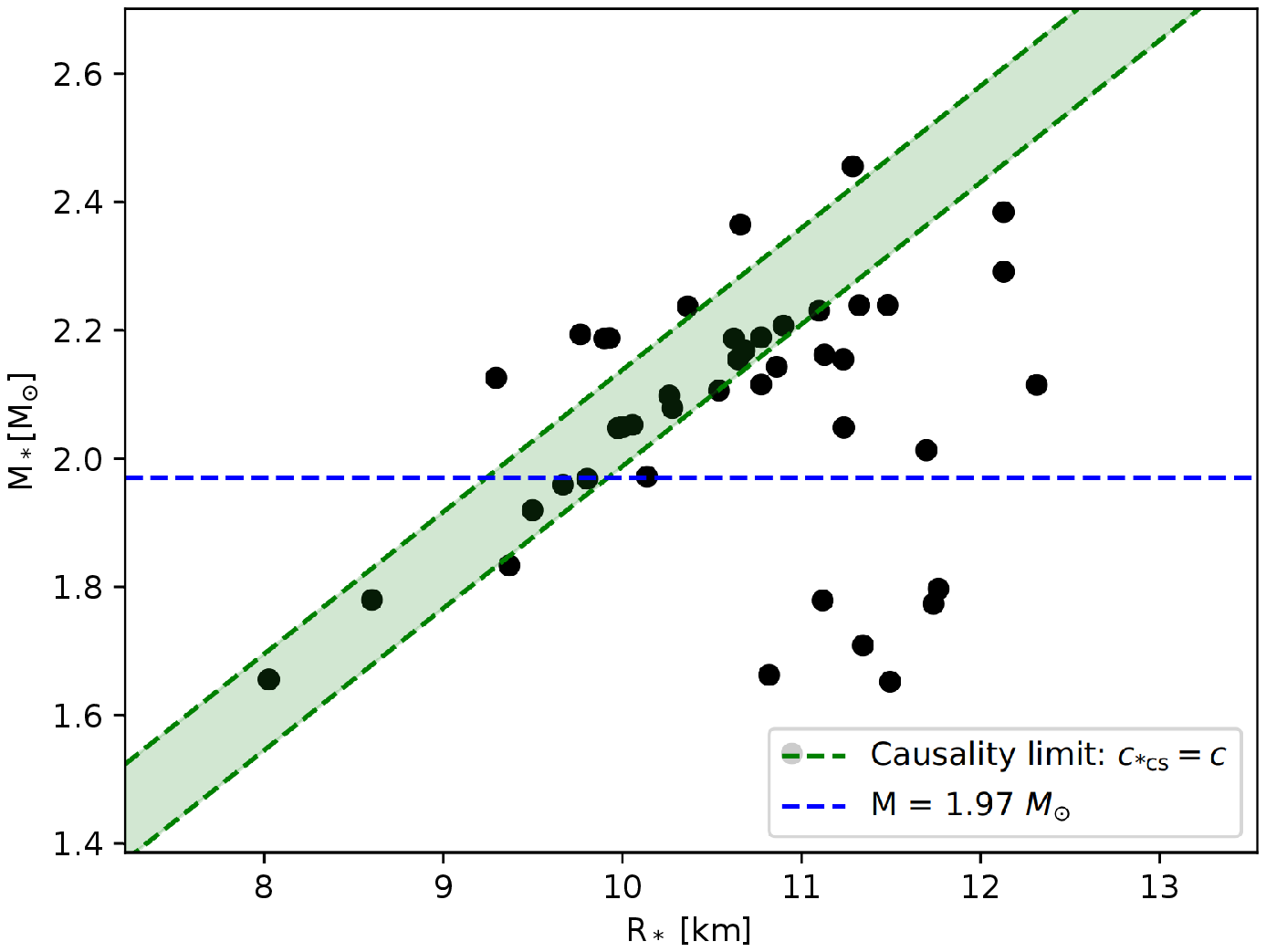}}\\
\subfloat[$\beta = -5$ restrictions in $P(\rho)$ plane]{\includegraphics[width = 9cm]{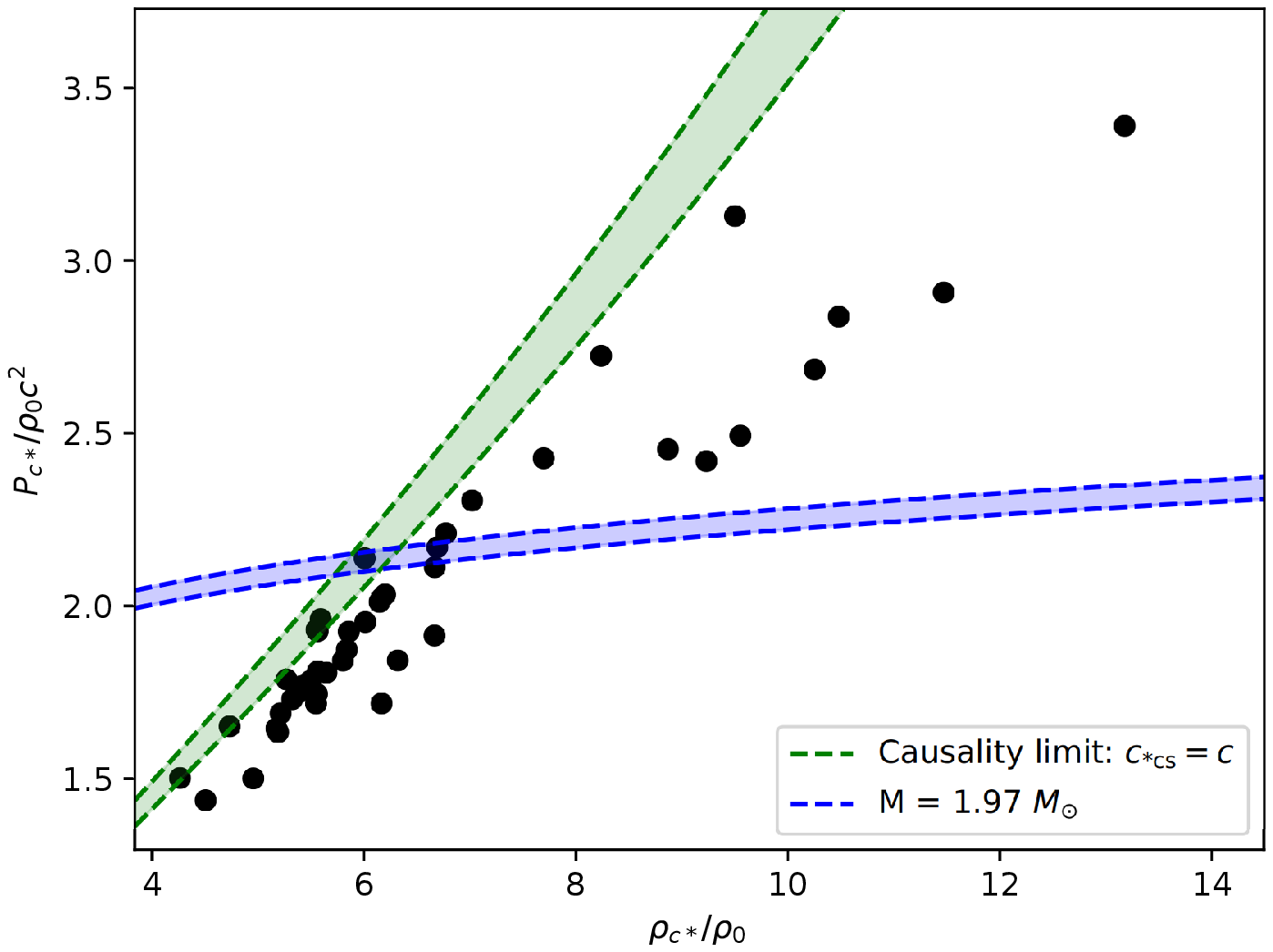}}
\subfloat[$\beta = -5$ restrictions in $M(R)$ plane]{\includegraphics[width = 9cm]{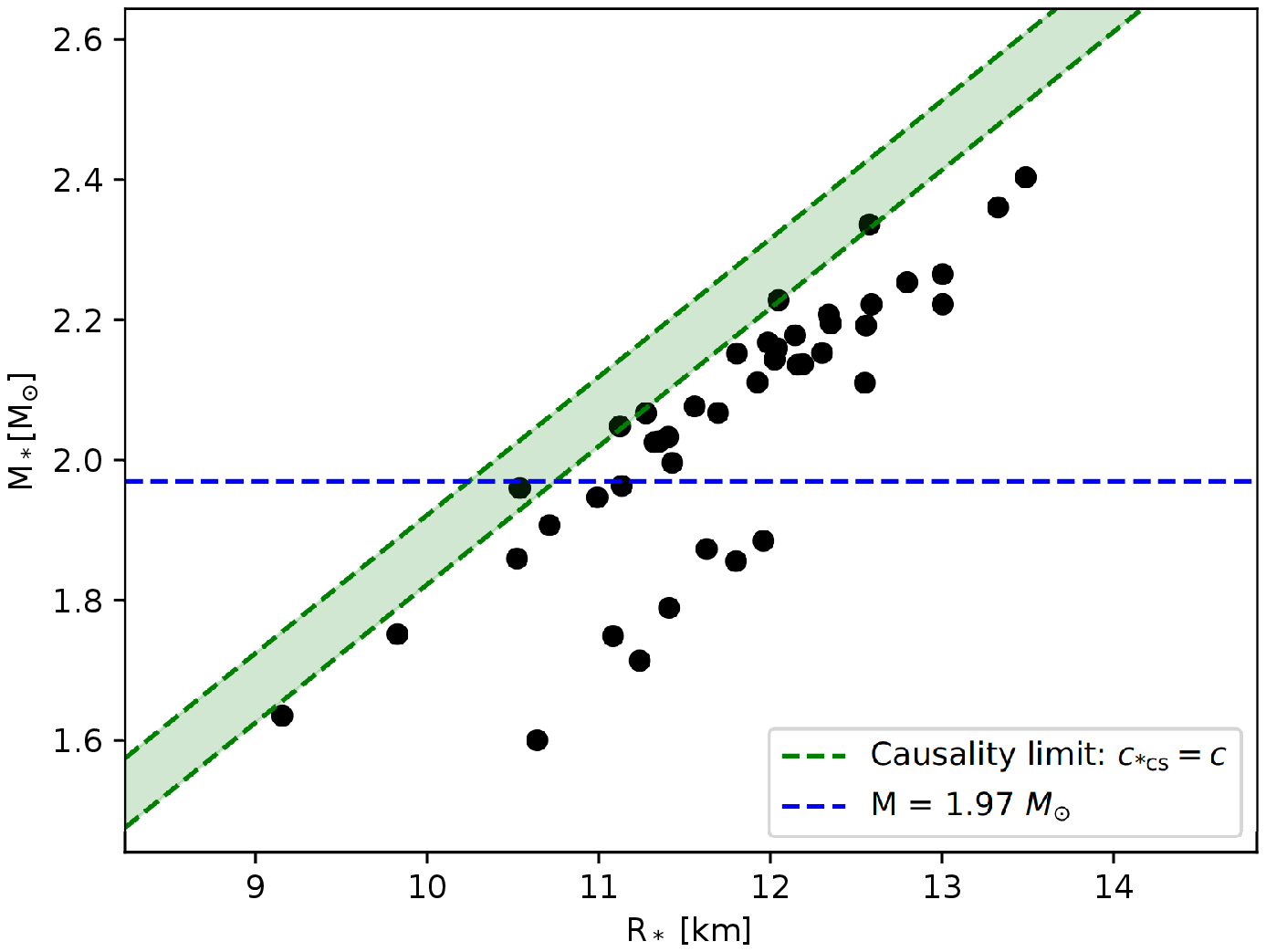}}\\
\caption{Final constraints on Mass-Radius and Pressure-Density using only baryonic EOS.}\label{Constraints_plot}
\end{figure*}

All restrictions resulting from each of the 6 theories with different $\beta$ have been obtained in both the Mass-Radius and Density-Pressure planes.
As becomes apparent from the expressions, larger fit error leads to significant broadening of the restrictions posed by the two conditions.
Indeed, the fits containing just 7 non-baryonic EOS fail to provide significant restrictions even in the GR case and so we have used the more severe restrictions posed by the baryonic-only fits similar to \cite{Ofengeim_2020}.
The effects obtained for GR and $\beta=-5$ are shown on Fig. \ref{Constraints_plot}

The general effect of the scalar-tensor theory can be broken down intuitively in two components.
First of all, the scalar field provides additional internal energy support against collapse which is proportional to the value of $\beta$ and supports higher values of the maximum mass with the increase of this parameter.
This naturally leads to the increase of maximum mass as was already observed on the Mass-Radius relation of Fig.\ref{ALF2_Example}.
The second consequence of this is that the maximum-mass point is reached at lower pressure values for higher $\beta$ values.
Therefore, even equations of state which violate causality in GR are reconciled with the constraint as early as $\beta = -5$.
Indeed, for values of $\beta > 6$ all considered EOS fit within the obtained constraints. 

\begin{figure}[!h]
\centering
\includegraphics[width = 9cm]{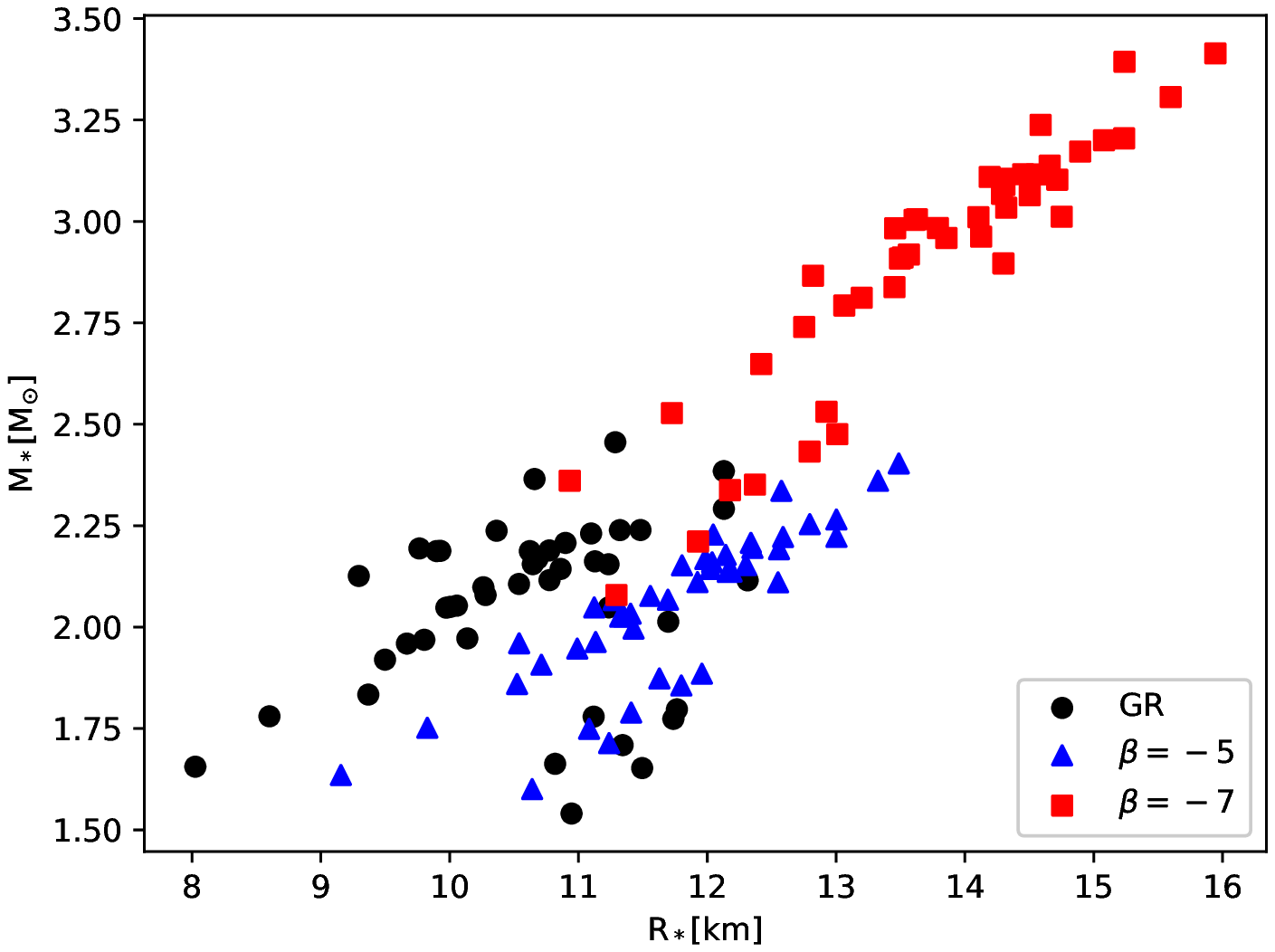}
\caption{Comparison between the maximum mass points plotted on the $M(R)$ plane for GR and scalar-tensor theory with $\beta$ values of $-5$ and $-7$ showing the reduced scatter of the points with reduced $\beta$.}\label{Lin_Scatter_Comparison}
\end{figure}

Interestingly, an additional effect of the scalar-tensor theory is to reduce the dispersion on the Mass--Radius relation, leading all EOS to approach linear dependence with the increase of the absolute value of $\beta$.
The general dependence between maximum mass and maximum radius for different EOS can be seen on Fig. \ref{Lin_Scatter_Comparison} for GR and the STT with $\beta = -5$ and $\beta = -7$.
The dependence looks linear for all theory parameters in a sense that stars with higher maximum radius tend to have higher maximum mass with straight proportionality.
It is evident from the plot, however, that the mean deviations from such a line become smaller with the increase of $\beta$'s absolute value which leads to a more visible structure in these cases.
No intuitive explanation for this result can be given by the authors at the current time. 

The results obtained can be used in two distinct ways.
Considering a fixed gravitational theory with known parameters (be it GR or a scalar-tensor theory with a particular value of $\beta$ and the considered $m_{\varphi} = 10^{-13}$ eV), we can constraint the EOS which lead to physically meaningful results and agree with current observational constraints.
These constraints can be quite different for the different gravitational theories.
On the other hand, the universality of the relations with respect to the EOS means that we can also make conclusions about the theories themselves and constraint their parameters based on measurements which are independent of the EOS.
The obvious preservation of the universality and high sensitivity of the fits to the theory parameters can prove to be a useful tool for the exploration of strong-regime gravity. For example it is possible that after taking into account other (not yet derived) constraints, a complete set of them will be incompatible for certain theories or values of the parameters.

\section{Conclusions}
In this paper we explored the universality (EOS independence) of relations connecting the neutron stars properties at the maximum-mass points for neutron star sequences in a certain class of massive scalar-tensor theories (STT).
We confirmed that this universality holds for a wide array of theory parameters with rms values at least as good as those for General Relativity.
More importantly, we discovered that the fit parameters of these universal relations vary significantly with the theory and its parameters. 
This provides a potential tool for constraining the different parameter values within a specific alternative theory of gravity in an EOS independent way. In addition, these relations can be used to judge how the currently available EOS constraints are altered with the inclusion of GR modifications.

We further studied the behavior of the fits under the addition of non-baryonic EOS, confirming that even with the addition of a small number of these, rms and maximum deviations of the fits are significantly worsened.
In order to keep the results unbiased, we have used a total of 53 equations of state including both polytropic approximations and tabulated versions of different high-density matter leading to a wide array of physical and even non-physical (from causality perspective) maximum mass states. 
Importantly, we have verified that the introduction of non-baryonic EOSs degrades the fit quality without changing the actual values for the fit parameters significantly.
On the other hand, STT leads to significantly different universal relation fit parameters compared to GR that are outside of the rms deviations even if non-baryonic EOS are taken into account.

Lastly, we derived constraints on the allowed EOS in the Mass-Radius and Pressure-Density planes based on observations and physical causality with the aid of the obtained universal relations.
In most cases, though, only the baryonic-only universal relations can produce good enough restrictions.
We demonstrated that these restrictions lead to the fact that more EOS are physically meaningful for higher absolute value of the $\beta$ parameter for the STT as compared to pure GR.
This methodology can be used in the following ways:
\begin{itemize}
\item{Assuming that a given gravitational theory is correct, the universal relations can be used to constrain the physical EOS based on different observations and theoretical limits, e.g. the NS maximum mass and causality. This is important  since currently most of the observations are interpreted with the assumption of GR as the correct theory that can cause difficulty when testing the strong field regime of gravity.}
\item{Assuming that the equation of state (EOS) is further constraint by the experiments, the relations can be used to constrain the allowed range of $\beta$ parameters for the considered class of STT. The novel experiments of heavy ion collisions promise to allow for such application within the coming decade.}
\item{If further independent constraints on the NS properties are obtained in the future, e.g. by the gravitational wave and electromagnetic observations, their simultaneous application can constraints the theory of gravity in an EOS independent way (e.g. it can happen that these constraints are incompatible for certain value of the STT parameters)}
\end{itemize}

\begin{acknowledgments}
We would like to thank Stoytcho S. Yazadjiev for the continuous support and guidance as well as many fruitful discussions in the preparation of this work.
We acknowledge the CompStar Online Suparnovae Equations of State (\href{https://compose.obspm.fr/home}{CompOSE}) for providing a large amount of freely available equations of state.
DD acknowledges financial support via an Emmy Noether Research Group funded by the German Research Foundation (DFG) under grant no. DO 1771/1-1. The Networking support
by the COST Actions CA16104 and CA16214 are also gratefully acknowledged.
\end{acknowledgments}

\section{Appendix}
We have listed all fitting coefficients in the following two Tables \ref{Primary_Fit_Table} and \ref{Secondary_Fit_Table}.

\begin{table*}
\centering
\begin{tabular}{||c|c|c c c c|c|c|c c c c|c|c||}
\cline{3-14}
\multicolumn{2}{c}{} & \multicolumn{6}{|c|}{All EoS} & \multicolumn{6}{c|}{Baryonic EoS} \\
\hline
Theory & $x_*$ & $a_x[\mathrm{km}]$ & $\varphi_x$ & $b_x[\mathrm{km}]$ & $\mathfrak{p}_x$ & rms & max & $a_x[\mathrm{km}]$ & $\varphi_x$ & $b_x[\mathrm{km}]$ & $\mathfrak{p}_x$ & rms & max  \\
\hline
\hline
\multirow{3}{*}{GR} & $P_{c*}$ & 9.3826 & -0.7576 & -3.4953 & 4.7933 & 0.0944 & 0.2507 & 9.1802 & -0.7402 & -2.7919 & 4.1425 & 0.0439 & 0.1129  \\
\cline{2-14}
& $\rho_{c*}$ & 41.2048 & 0.1932 & -14.7431 & 4.9210 & 0.0451 & 0.1601 & 35.5174 & 0.3091 & -5.2840 & 3.0607 & 0.0178 & 0.0399  \\
\cline{2-14}
& $c_{sc*}$ & 2.7995 & -0.9228 & -1.6469 & 1.1628 & 0.0823 & 0.2345 & 2.8984 & -0.9918 & -2.5275 & 1.1239 & 0.0290 & 0.0720  \\
\hline
\multirow{3}{*}{$\beta= -5$} & $P_{c*}$ & 7.6257 & 1.6200 & 0.0754 & 1.9942 & 0.0054 & 0.0156 & 8.0789 & 1.6229 & -0.4527 & 2.1846 & 0.0048 & 0.0143  \\
\cline{2-14}
& $\rho_{c*}$ & 8.3788 & 1.7998 & -0.8300 & 2.5683 & 0.0376 & 0.1033 & 11.8459 & 1.8330 & -5.8404 & 5.8969 & 0.0246 & 0.0788 \\
\cline{2-14}
& $c_{sc*}$ & 0.3970 & -1.0631 & -0.3816 & 0.3919 & 0.0648 & 0.2040 & 0.8322 & -1.0435 & -0.5843 & 0.6055 & 0.0377 & 0.0917  \\
\hline
\multirow{3}{*}{$\beta= -5.5$} & $P_{c*}$ & 7.2364 & 1.7198 & -0.2244 & 2.1273 & 0.0099 & 0.0332 & 8.1055 & 1.7333 & -1.4188 & 2.6400 & 0.0078 & 0.0291  \\
\cline{2-14}
& $\rho_{c*}$ & 8.9191 & 1.8365 & -1.7971 & 3.2847 & 0.0417 & 0.1280 & 10.8269 & 1.8727 & -4.9321 & 5.4391 & 0.0255 & 0.0829  \\
\cline{2-14}
& $c_{sc*}$ & 0.4442 & -1.0461 & -0.4506 & 0.3879 & 0.0598 & 0.1923 & 0.9148 & -1.0277 & -0.6869 & 0.6005 & 0.0337 & 0.0921  \\
\hline
\multirow{3}{*}{$\beta= -6$} & $P_{c*}$ & 7.2268 & 1.7765 & -0.3837 & 2.2448 & 0.0182 & 0.0486 & 7.8176 & 1.8014 & -1.5253 & 2.7439 & 0.0121 & 0.0391  \\
\cline{2-14}
& $\rho_{c*}$ & 9.4505 & 1.8624 & -2.2587 & 3.6574 & 0.0471 & 0.1565 & 9.6524 & 1.9031 & -3.5436 & 4.4582 & 0.0269 & 0.0872  \\
\cline{2-14}
& $c_{sc*}$ & 0.4840 & -1.0287 & -0.5193 & 0.3809 & 0.0571 & 0.1832 & 0.9934 & -1.0107 & -0.7840 & 0.5945 & 0.0317 & 0.0942  \\
\hline
\multirow{3}{*}{$\beta= -6.5$} & $P_{c*}$ & 7.6620 & 1.8204 & -0.8942 & 2.5374 & 0.0239 & 0.0736 & 7.8476 & 1.8519 & -1.7477 & 2.8874 & 0.0144 & 0.0462  \\
\cline{2-14}
& $\rho_{c*}$ & 10.2557 & 1.8875 & -2.9615 & 4.1469 & 0.0500 & 0.1750 & 9.4400 & 1.9300 & -3.0904 & 4.1035 & 0.0270 & 0.0867  \\
\cline{2-14}
& $c_{sc*}$ & 0.5511 & -1.0103 & -0.6140 & 0.3868 & 0.0549 & 0.1746 & 1.1137 & -0.9923 & -0.9133 & 0.6052 & 0.0304 & 0.0926  \\
\hline
\multirow{3}{*}{$\beta= -7$} & $P_{c*}$ & 8.5373 & 1.8560 & -1.7660 & 3.0279 & 0.0295 & 0.0987 & 8.3905 & 1.8901 & -2.3423 & 3.2404 & 0.0173 & 0.0544  \\
\cline{2-14}
& $\rho_{c*}$ & 11.5900 & 1.9108 & -4.1714 & 4.9313 & 0.0528 & 0.1924 & 9.9254 & 1.9535 & -3.3686 & 4.2699 & 0.0280 & 0.0833  \\
\cline{2-14}
& $c_{sc*}$ & 0.6032 & -0.9927 & -0.7103 & 0.3853 & 0.0532 & 0.1677 & 1.2056 & -0.9748 & -1.0351 & 0.6020 & 0.0298 & 0.0915  \\
\hline
\end{tabular}
\caption{Primary fit parameters for General Relativity and the Scalar-Tensor theories}\label{Primary_Fit_Table}
\end{table*}

\begin{table*}
\centering
\begin{tabular}{||c|c|c c c c|c|c|c c c c|c|c||}
\cline{3-14}
\multicolumn{2}{c}{} & \multicolumn{6}{|c|}{All EoS} & \multicolumn{6}{c|}{Baryonic EoS} \\
\hline
Theory & $y_*$ & $A_y$ & $\Phi_x$ & $B_y$ & $q_y$ & rms & max & $A_y$ & $\Phi_x$ & $B_y$ & $q_y$ & rms & max  \\
\hline
\hline
\multirow{3}{*}{GR} & $M_*$ & 38.4810 & -0.2984 & 0.2325 & 1.000 & 0.0173 & 0.0560 & 31.8824 & -0.3219 & 0.04492 & 1.000 & 0.0083 & 0.0223  \\
\cline{2-14}
& $R_*$ & 34.8616 & 0.06353 & 0.3807 & 1.000 & 0.0198 & 0.0780 & 27.8131 & 0.1161 & 0.1830 & 1.000 & 0.0091 & 0.0228  \\
\cline{2-14}
& $c_{sc*}$ & -4.7464 & -0.8937 & 0.0578 & -1.1628 & 0.0576 & 0.2546 & -4.9774 & -0.7857 & 0.0652 & -1.1239 & 0.0180 & 0.0669  \\
\hline
\multirow{3}{*}{$\beta= -5$} & $M_*$ & 9.9477 & 1.8043 & -0.0325 & 1.000 & 0.0032 & 0.0139 & 10.4709 & 1.8575 & -0.0857 & 1.000 & 0.0021 & 0.0073  \\
\cline{2-14}
& $R_*$ & -62.5401 & -0.7258 & 0.0806 & 1.000 & 0.0270 & 0.0897 & -67.9677 & -0.5831 & 0.3807  & 1.000 & 0.0124 & 0.0387  \\
\cline{2-14}
& $c_{sc*}$ & -59.9871 & -0.5147 & 0.1071 & -0.3919 & 0.0673 & 0.2470 & -33.5269 & -0.3724 & 0.4598 & -0.6055 & 0.0186 & 0.0771  \\
\hline
\multirow{3}{*}{$\beta= -5.5$} & $M_*$ & 19.8836 & 2.1793 & -0.0897 & 1.000 & 0.0064 & 0.0239 & 21.4495 & 2.1994 & -0.1263 & 1.000 & 0.0020 & 0.0046  \\
\cline{2-14}
& $R_*$ & -96.5724 & -0.6696 & 0.1387 & 1.000 & 0.0318 & 0.1047 & -94.9658 & -0.6269 & 0.2662 & 1.000 & 0.0117 & 0.0347  \\
\cline{2-14}
& $c_{sc*}$ & -72.6784 & -0.5152 & 0.1740 & -0.3879 & 0.0687 & 0.2747 & -35.1266 & -0.4487 & 0.3559 & -0.6005 & 0.0144 & 0.0600  \\
\hline
\multirow{3}{*}{$\beta= -6$} & $M_*$ & 33.0359 & 2.3194 & -0.1238 & 1.000 & 0.0082 & 0.0254 & 33.2303 & 2.2841 & -0.0927 & 1.000 & 0.0024 & 0.0049  \\
\cline{2-14}
& $R_*$ & -134.5865 & -0.6423 & 0.1596 & 1.000 & 0.0338 & 0.1079 & -117.8133 & -0.6664 & 0.1680 & 1.000 & 0.0112 & 0.0295  \\
\cline{2-14}
& $c_{sc*}$ & -86.5895 & -0.5212 & 0.1932 & -0.3809 & 0.0696 & 0.2729 & -35.5925 & -0.5146 & 0.2457 & -0.5945 & 0.0130 & 0.0536  \\
\hline
\multirow{3}{*}{$\beta= -6.5$} & $M_*$ & 51.9580 & 2.3862 & -0.1302 & 1.000 & 0.0093 & 0.0289 & 48.7508 & 2.3302 & -0.0639 & 1.000 & 0.0029 & 0.0082  \\
\cline{2-14}
& $R_*$ & -183.9125 & -0.6323 & 0.1640 & 1.000 & 0.0348 & 0.1117 & -149.5907 & -0.6808 & 0.1142 & 1.000 & 0.0113 & 0.0337  \\
\cline{2-14}
& $c_{sc*}$ & -98.6345 & -0.5327 & 0.2008 & -0.3868 & 0.0648 & 0.2418 & -37.4032 & -0.5529 & 0.1836 & -0.6052 & 0.0123 & 0.0507  \\
\hline
\multirow{3}{*}{$\beta= -7$} & $M_*$ & 79.1012 & 2.4239 & -0.1351 & 1.000 & 0.0098 & 0.0301 & 69.7771 & 2.3534 & -0.0414 & 1.000 & 0.0031 & 0.0103  \\
\cline{2-14}
& $R_*$ & -250.8564 & -0.6264 & 0.1703 & 1.000 & 0.0343 & 0.1122 & -192.9997 & -0.6904 & 0.0839 & 1.000 & 0.0106 & 0.0354  \\
\cline{2-14}
& $c_{sc*}$ & -118.3153 & -0.5418 & 0.2117 & -0.3853 & 0.0580 & 0.1832 & -42.4449 & -0.5799 & 0.1511 & -0.6020 & 0.0116 & 0.0492  \\
\hline

\end{tabular}
\caption{Secondary fit parameters for General Relativity and the Scalar-Tensor theories}\label{Secondary_Fit_Table}
\end{table*}

\end{document}